\documentclass[conference, a4paper]{IEEEtran}

\IEEEoverridecommandlockouts                            
\usepackage{booktabs}
\usepackage{graphicx}

\usepackage{multirow, makecell}

\usepackage{amsmath,amssymb,latexsym}
\usepackage{rotating}
\usepackage{lettrine}
\usepackage{textcomp}
\usepackage[table]{xcolor}
\usepackage{alltt}
\usepackage{hyperref}
\usepackage{siunitx}
\usepackage{tikz}

\usepackage{xcolor}

\usepackage{flushend}

\usepackage{lipsum}
\usepackage[pscoord]{eso-pic}
\newcommand{\placetextbox}[3]{
  \setbox0=\hbox{#3}
  \AddToShipoutPictureFG*{
    \put(\LenToUnit{#1\paperwidth},\LenToUnit{#2\paperheight}){\vtop{{\null}\makebox[0pt][c]{#3}}}%
  }%
}%
\newcommand{\mymargin}{0.0cm}

\title{Machine Learning to Predict Slot Usage in TSCH Wireless Sensor Networks
\thanks{This work was partially supported by the European Union under the Italian National Recovery and Resilience Plan (NRRP) of NextGenerationEU, partnership on ``Telecommunications of the Future'' (PE00000001 - program ``RESTART'').}
}

\author{
    \IEEEauthorblockN{
    Stefano Scanzio\IEEEauthorrefmark{1},
    Gabriele Formis\IEEEauthorrefmark{1}\IEEEauthorrefmark{2},
    Tullio Facchinetti\IEEEauthorrefmark{3},
    Gianluca Cena\IEEEauthorrefmark{1}
    }
    
    \IEEEauthorblockA{\IEEEauthorrefmark{1}National Research Council of Italy (CNR--IEIIT), Italy.}    
    
    \IEEEauthorblockA{\IEEEauthorrefmark{2}Politecnico di Torino, Italy.}
    \IEEEauthorblockA{\IEEEauthorrefmark{3}University of Pavia, Italy.}
    Emails: stefano.scanzio@cnr.it, gabriele.formis@polito.it, tullio.facchinetti@unipv.it,\\ gianluca.cena@cnr.it
    }

\begin{document}
\placetextbox{0.5}{1}{This is the author's version of an article that has been published.}
\placetextbox{0.5}{0.985}{Changes were made to this version by the publisher prior to publication.}
\placetextbox{0.5}{0.97}{The final version of record is available at \href{https://doi.org/10.1109/ETFA65518.2025.11205770}{https://doi.org/10.1109/ETFA65518.2025.11205770}}%
\placetextbox{0.5}{0.05}{Copyright (c) 2025 IEEE. Personal use is permitted.}
\placetextbox{0.5}{0.035}{For any other purposes, permission must be obtained from the IEEE by emailing pubs-permissions@ieee.org.}%

\maketitle
\thispagestyle{empty}
\pagestyle{empty}

\begin{abstract}
Wireless sensor networks (WSNs) are employed across a wide range of industrial applications where ultra-low power consumption is a critical prerequisite. At the same time, these systems must maintain a certain level of determinism to ensure reliable and predictable operation. In this view, time slotted channel hopping (TSCH) is a communication technology that meets both conditions, making it an attractive option for its usage in industrial WSNs.

This work proposes the use of machine learning to learn the traffic pattern generated in networks based on the TSCH protocol, in order to turn nodes into a deep sleep state when no transmission is planned and thus to improve the energy efficiency of the WSN. The ability of machine learning models to make good predictions at different network levels in a typical tree network topology was analyzed in depth, showing how their capabilities degrade while approaching the root of the tree. The application of these models on simulated data based on an accurate modeling of wireless sensor nodes indicates that the investigated algorithms can be suitably used to further and substantially reduce the power consumption of a TSCH network.
\end{abstract}



\section{Introduction}
\label{sec:introduction}

In the context of wireless networks, time division multiple access (TDMA) is a communication paradigm based on the slicing of time into slots of fixed duration. The TDMA concept is the main building block for enabling scheduling in such networks, providing the double advantages of increased determinism and reduced power consumption. The latter is obtained by turning the node into a deep sleep state when no transmission is scheduled for it in the considered time slot. Grounded on the concept of TDMA, the time slotted channel hopping (TSCH) operating mode described in the IEEE 802.15.4 standard~\cite{IEEE-802.15.4-2020} operates as TDMA in terms of time division into slots (time slotted), while adding the possibility of changing the transmission channel at every transmission, reducing the variability in performance due to the different quality of diverse channels (channel hopping)~\cite{10158374}.

At the same time, wireless sensor networks (WSNs) are gaining ground in the landscape of industrial networks, which are commonly characterized by a high level of heterogeneity in terms of communication technologies~\cite{SCANZIO2021103388}.
WSNs are used in all these industrial contexts, at the edge of the network~\cite{https://doi.org/10.1002/ett.70088}, where sensing or actuation are not limited to strict timing constraints (hard deadlines), but on the contrary, low power consumption is the main target. This is common in applications such as environmental sensing for Heating, Ventilation and Air Conditioning (HVAC) systems, in product tracking, for predictive maintenance, in backup networks, or alarm systems~\cite{majid2022applications}. In all these applications, the WSN follows a paradigm called \textit{deploy and forget}, where battery-powered nodes are permanently placed in the environment under monitoring and must operate without external intervention for 10 years or more.

Among the several different protocols used for implementing WSNs (e.g., Zigbee, WirelessHART, ISA 100.11a, LoRaWAN, IO-Link Wireless, or Bluetooth Low Energy), in the context of industrial networks, TSCH has the dual advantage of providing a good level of determinism~\cite{CENA2020102199} and an energy consumption that is low enough to be suitable for the \textit{deploy and forget} paradigm. Both properties derive from the time slotted characteristic of TSCH, which enables proper scheduling of network traffic~\cite{BALBI2025104164}. In particular, when no transmission is scheduled for a given node in a specific time slot, the node can enter a deep sleep state, causing the node to consume a negligible amount of energy. 

Like many WSN implementations, common TSCH networks are based on a tree routing topology, which is dynamically constructed by the routing protocol for low-power and lossy networks (RPL)~\cite{ZIRAK2025103843}. In this case, values sensed and generated by leaf nodes are delivered through a multi-hop path to the sink node, i.e., the root of the tree.

On a single link of the tree, transmissions can occur only in given time intervals/slots as predefined by a schedule. Since typically these networks make use of a small subset of the scheduled slots, specific algorithms can be used to predict which slots are effectively used for transmissions. For unused slots, the node can enter a deep sleep state to significantly reduce its power consumption.

This work investigates the application of machine learning (ML) approaches to the prediction of the patterns generated in a TSCH network.

Due to the tree topology of the network, moving from the leaves to the root node, the traffic flows generated by leaf nodes aggregate; this reduces the capability of predicting the used slots. One goal of this work is to examine how the performance of ML algorithms for the prediction of slot utilization degrades while approaching the root node. On the one hand, this performance degradation is a very important aspect that highlights one of the possible limitations of a data-driven approach to the addressed task. On the other hand, results demonstrate that the performance remains satisfactory. The analysis is done with energy savings in mind, for which the results reported in this paper show substantial reductions.

The paper is organized as follows. Section~\ref{sec:review} analyzes the scientific literature about techniques to reduce power consumption in TSCH networks, while Section~\ref{sec:tsch} reports the basic concepts behind TSCH and highlights some aspects related to power consumption. Section~\ref{sec:algo} describes the mathematical framework for the prediction of slot utilization, the related results are commented in Section~\ref{sec:results}, and Section~\ref{sec:conclusions} finally concludes this work.

\section{Literature Review}
\label{sec:review}

The research about reduction in energy consumption for WSN in general, and TSCH specifically, has been and is currently a very active area and direction of investigation.

Some researches involve all the nodes of the network as a whole, and they are based on global optimizations aimed at enhancing some performance indicators such as power consumption. In this context, a common strategy is to balance nodes' load to have a uniform discharge of nodes' batteries~\cite{doi:10.1504/IJAHUC.2020.107505}, or the tuning of specific network parameters to improve one or more key performance indicators such as latency, reliability, or power consumption~\cite{9187609}. It is important to notice that, typically, the optimization is subject to trade-offs; in other words, the improvement of one index leads to the worsening of the others.

Other techniques optimize specific protocols that are used in combination with TSCH to setup and manage the network. For instance, in~\cite{9058396}, a more energy-efficient version of the RPL protocol, which is aimed at generating the network routing topology, is presented. In~\cite{MOHAMADI20221}, the authors proposed an energy-efficient network formation mechanism.

In order to divide time into slots, nodes must operate in a synchronized manner~\cite{MONGELLI20161}. The synchronization accuracy and efficiency of the protocol have a direct impact on power consumption. In~\cite{2019_dynamic_guard}, the guard time, which is used to make the protocol robust to not perfect synchronization, is adapted dynamically, while in~\cite{s19194128} a dynamic modification of the sending period of enhanced beacons is proposed, which are directly used by the TSCH protocol for synchronization purposes.

Techniques based on white and black listing permit the selection of the best communication channel or the removal of the worst channel in order to improve communication quality~\cite{2022-Electronics-Listing, 9925697}. Since this also reduces the number of retransmissions, it has a direct impact on energy consumption.

A very common way to reduce power consumption is by acting on scheduling properties. Following this approach, there are methods, such as the VAM-HSA algorithm~\cite{7962719}, which reduce energy consumption by solving an energy-efficiency maximization problem, while~\cite{8371162} proposed a schedule procedure named PRCOS to maximize system lifetime. The method proposed in~\cite{10770109} extends the concept of scheduling to reduce energy consumption for battery-less devices.

Finally, proactive reduction of idle listening (PRIL) techniques take advantage of the periodicity of traffic to place nodes in deep sleep states when a scheduled slot is not used. In this context, PRIL-F \cite{10.1007/978-3-030-61746-2_11,9903301} was the first definition of the technique that acts only one hop away from the source of the traffic, while PRIL-M \cite{10540576} overcomes this limitation.
The PRIL techniques, even if quite different, are the most similar to the concepts described in this paper, but they do not rely on self-learning. Indeed, PRIL techniques make use of additional information added to packets for entering nodes into deep sleep states, whereas it is not the case for the ML technique presented in this work.

\section{TSCH and Power Consumption}
\label{sec:tsch}
\begin{figure}[t]
\begin{center}
\includegraphics[width=1.0\linewidth]{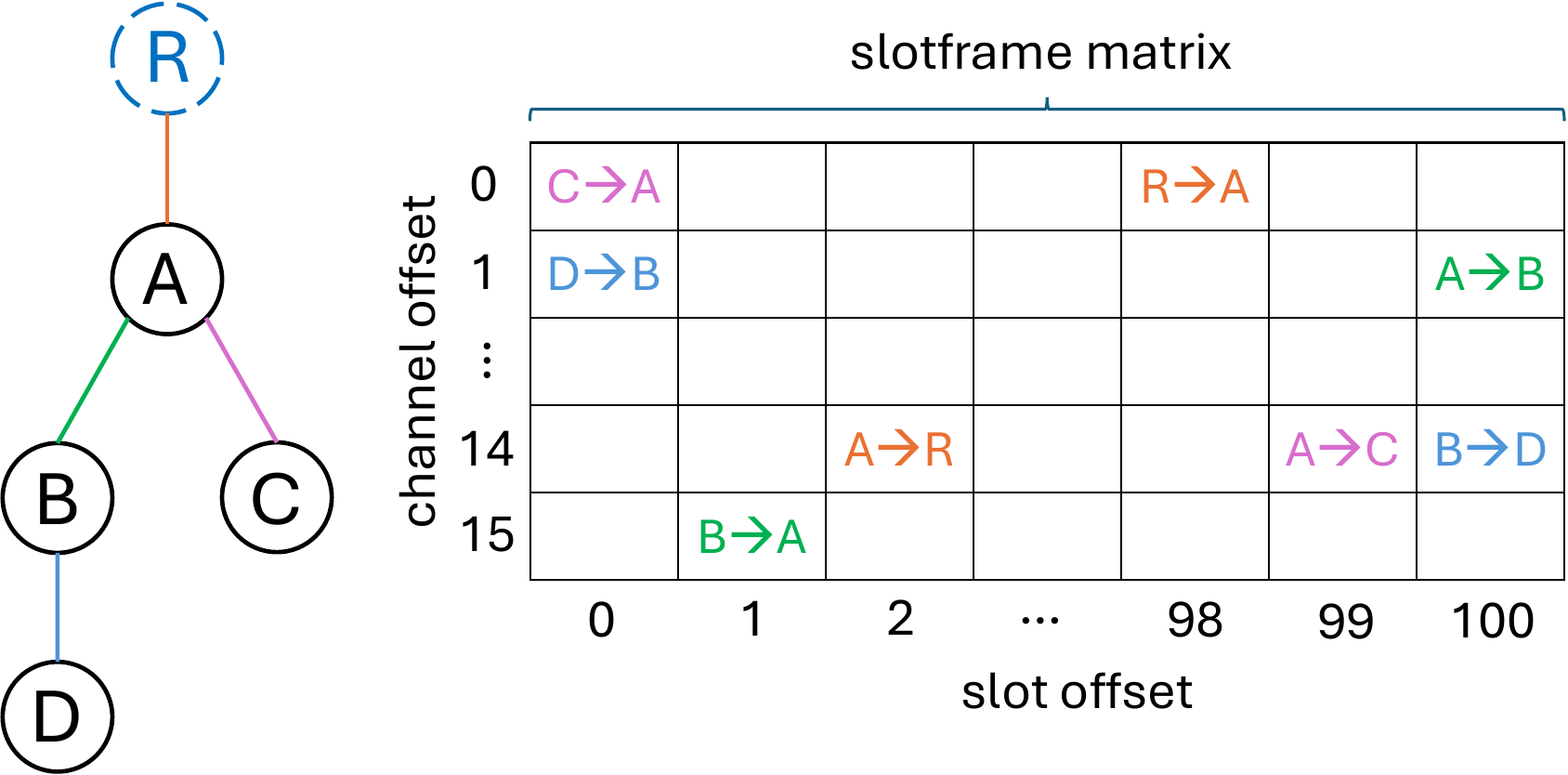}
\end{center}
\caption{Example of a TSCH slotframe matrix.}
\label{fig:matrix}
\end{figure}

The TSCH protocol technology was first standardized in 2012, and it is currently included in the latest version of the IEEE 802.15.4 standard~\cite{IEEE-802.15.4-2020}. Unlike the deterministic and synchronous multi-channel extension (DSME) operating mode, in TSCH time is slotted similarly to TDMA. This featureincreases latency predictability~\cite{9187609,CENA2020102199,Shudrenko2024}. The TSCH operating mode defines the operation of a WSN network at the MAC layer. Protocol suites such as IPv6 over the TSCH mode of IEEE 802.15.4e (6TiSCH)~\cite{2021-IETF-6TiSCH,8823863} embed and describe the interaction of a set of protocols in addition to TSCH (e.g., RPL to define routing) to provide the user with all the required components to setup a TSCH network. In addition to 6TiSCH, other protocols such as WirelessHART and ISA100.11a are based on the concepts of TSCH.

In TSCH, time is divided into \textit{slots} (or \textit{timeslots}) of fixed duration $T_{\mathrm{slot}}$ (Fig.~\ref{fig:matrix}). Transmissions, which can be acknowledged or not, have to be concluded within the slot. In real implementations, the slot has a duration of \SI{10}{} or \SI{20}{ms}.
Without loss of generality, the analysis carried out in this work concentrates only on acknowledged transmissions, which are those most commonly used in the industrial context.

A schedule can be defined to reserve a specific transmission opportunity between a source and a destination node. In the TSCH terminology, a \textit{link} represents the slot reserved for the transmission between a sender and a receiver.
In particular, in TSCH the \textit{slotframe matrix} defines the schedule, i.e., the set of links. Each node has its own copy of the TSCH matrix, in which only the information pertaining to the incoming and outgoing transmissions regarding this specific node is reported.

The \textit{slot offset} represents the position in terms of column in the TSCH slotframe matrix. Instead, the rows of the matrix are identified by the \textit{channel offset}. In particular, the number of rows of the matrix is equal to the number of physical channels. In the common case of a WSN operating on the \SI{2.4}{GHz} band, the standard defines 16 channels, each spaced by \SI{5}{MHz} and a channel bandwidth of \SI{3}{MHz}. In particular, the first channel 11 is centered at the frequency \SI{2.405}{GHz}, while the last channel 26 is centered at the frequency \SI{2.490}{GHz}.

When a TSCH network is switched on, the absolute slot number (ASN) counter is initialized and incremented at every slot. This counter is kept synchronized on all nodes in the network and can be used as a reference time to perform the operations defined by the schedule synchronously. Given a link in the slotframe matrix, which is identified by the couple slot and channel offset, the ASN is used to obtain the real physical channel of transmission using the following formula:

\begin{equation}
\mathrm{ch}_{\mathrm{phy}} = \mathrm{H}[ (\mathrm{ASN} + \mathrm{ch}_{\mathrm{offset}})\ \mathrm{mod}\ |\mathrm{H}| ],
\label{eq:channel_hopping}
\end{equation}\smallskip

where $\mathrm{ch}_{\mathrm{phy}}$ is the physical channel, $\mathrm{ch}_{\mathrm{offset}}$ is the channel offset, $\mathrm{H}[\cdot]$ is the hop sequence list, i.e., a pseudo-random sequence of channels used to associate a position in the list given by $(\mathrm{ASN} + \mathrm{ch}_{\mathrm{offset}})\ \mathrm{mod}\ |\mathrm{H}|$ with the physical channel, and $|\mathrm{H}|$ is its length. Typically, $|\mathrm{H}|$ is equal to the number of physical channels.
In the experimental analysis reported in this work, it is assumed $|\mathrm{H}|=16$.

\begin{figure}[t]
\begin{center}
\includegraphics[width=1\linewidth]{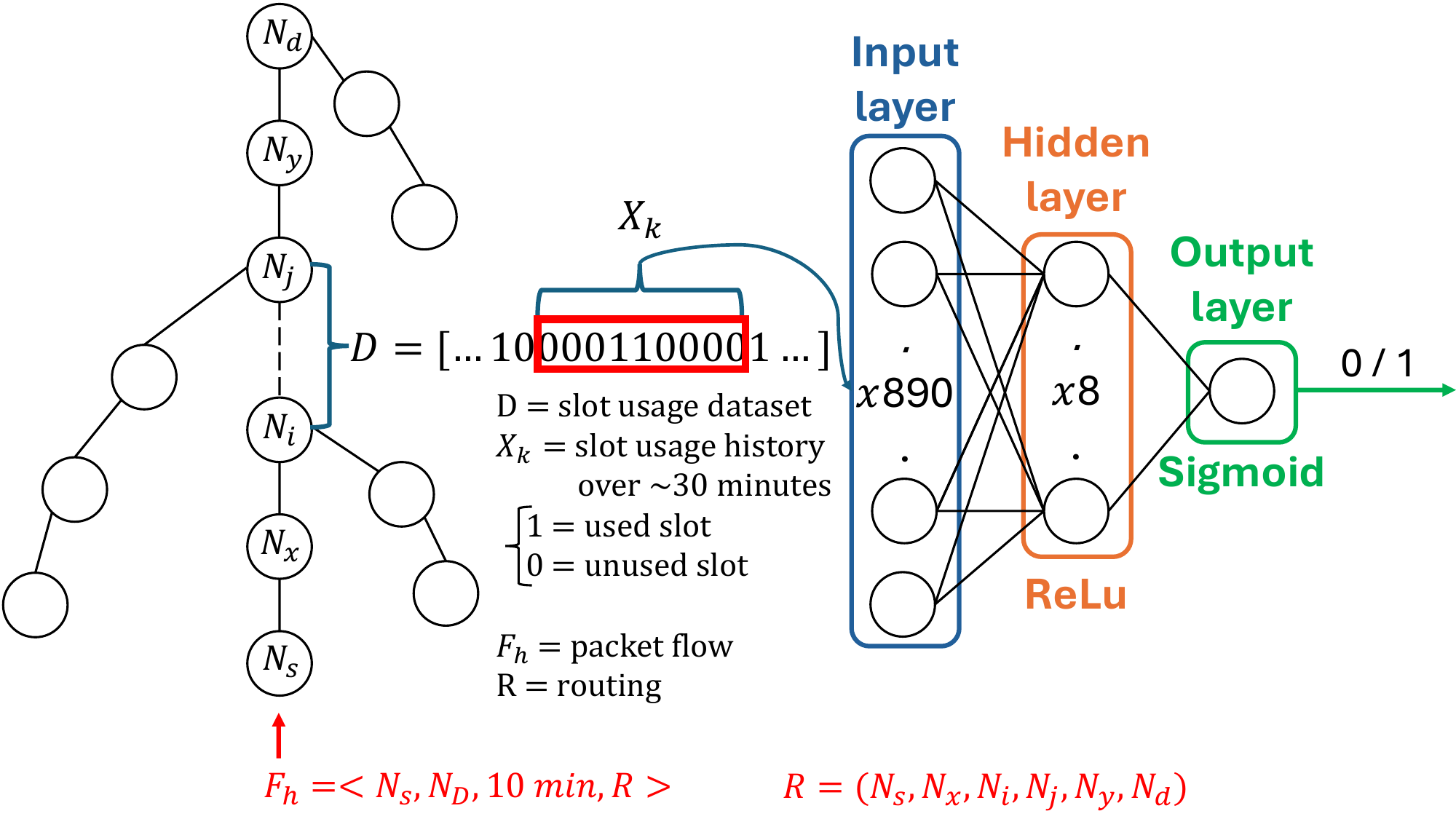}
\end{center}
\caption{Example network topology that shows a routing and the corresponding representation.\label{fig:ex_net}}
\end{figure}

While the number of rows of the slotframe matrix is identified by the number of physical channels, the number of columns $N_{\mathrm{slot}}$ is configurable, and a common value is $N_{\mathrm{slot}}=101$. The schedule reported in the matrix repeats over time.
As a consequence, every $T_{\mathrm{matrix}}=N_{\mathrm{slot}}\cdot T_{\mathrm{slot}}$, a scheduled link is repeated.
An example of TSCH slotframe matrix is reported in Fig.~\ref{fig:matrix}.
In the example, the value of $T_{\mathrm{matrix}}$ is \SI{2.02}{s}.

There are two key points to consider at this stage. The first regards the characteristic of the protocol called channel hopping. In the case the transmission of a frame does not succeed (i.e., the acknowledge (ACK) frame is not received by the sender), it is retransmitted in the next available link between the nodes involved in the exchange. If only a link is scheduled between these nodes in the slotframe matrix, this happens after $T_{\mathrm{matrix}}=\SI{2.02}{s}$. Due to (\ref{eq:channel_hopping}), each time a retransmission takes place, a new physical channel is computed, which typically differs from the channel used for the previous transmission. Channel hopping permits the perception of a channel between the two nodes with less variability in terms of frame delivery ratio, since both the worst and the best channels are selected probabilistically with a probability equal to that of the other channels.

The second point is more related to power consumption. From an application perspective, the traffic flows generated by nodes in WSNs generally have periods in the range of minutes. This means that, with a $T_{\mathrm{matrix}}=\SI{2.02}{s}$, most of the scheduled links are not used. For example, without taking into account retransmissions, if the generation period of a periodic flow is \SI{10}{minutes}, only $1$ slot is actually used in every $297$ slots. For the other slots, the receiver node turns on its reception circuitry to receive frames, but it wastes energy since no frame is actually sent. This phenomenon, which is known as \textit{idle listening}, is very impactful for this kind of network since idle listening consumes about half the energy used to actually receive a frame~\cite{9187609,2014-SJ-consumption}.

\section{Slot usage prediction}
\label{sec:algo}

Let $N_i\!\rightarrow\!N_j$ be a scheduled link of a transmission between node $N_i$ and $N_j$. The notation $N_i\!\xrightarrow{l}\!N_j$ identifies a link that is placed at level $l$ of the network, where $l=1$ identifies the lowest level between a leaf node and a node of the next level, $l=2$ is the following upper level, and so on.

As is common in industrial networks, packet flows are periodic. A specific flow $\mathcal{F}_h$ is defined by a tuple $\langle N_s, N_d, P, R \rangle$, where $N_s$ is the source node, $N_d$ is the destination node, $P$ is the period, and $R$ is the routing, i.e., an ordered sequence representing the nodes that need to be traversed to reach the destination.
In the case of the example of Fig.~\ref{fig:ex_net}, $R=(N_s, N_x, N_i, N_j, N_y, N_d)$, and involved links with corresponding levels are $N_s\!\xrightarrow{l=1}\!N_x$, $N_x\!\xrightarrow{l=2}\!N_i$,
$N_i\!\xrightarrow{l=3}\!N_j$,
$N_j\!\xrightarrow{l=4}\!N_y$,
$N_y\!\xrightarrow{l=5}\!N_d$.

Let $x^{N_i\!\rightarrow\!N_j}_k=1$ denote if a given link ${N_i\!\rightarrow\!N_j}$ is used (by flows crossing the link) for transmission at a discrete time $k$, while $x^{N_i\!\rightarrow\!N_j}_k=0$ denotes if the link is not used. As already introduced, this latter condition is known as idle listening, a state in which the receiver node $N_j$ activates its interface to receive a frame that will not arrive, resulting in a significant waste of energy. It is important to notice that, if the TSCH matrix contains only one scheduled link $N_i\!\rightarrow\!N_j$ between these two nodes, a value $x^{N_i\!\rightarrow\!N_j}_k$ is logged for every repetition of the slotframe matrix, which typical period is $T_{\mathrm{matrix}}=$\SI{2.02}{s}. If the same link $N_i\!\rightarrow\!N_j$ is present more than once in the TSCH matrix, it is counted according to the number of occurrences (i.e., each occurrence corresponds to the acquisition of a sample $x^{N_i\!\rightarrow\!N_j}_k$).

The ordered sequence of acquired samples $(x^{N_i\!\rightarrow\!N_j}_1,...,x^{N_i\!\rightarrow\!N_j}_k,...,x^{N_i\!\rightarrow\!N_j}_N)$ is the dataset $\mathcal{D}^{N_i\!\rightarrow\!N_j}$ with cardinality $N$, which represents the slot usage for the link $N_i\!\rightarrow\!N_j$. To simplify the notation, when referring to a specific link, the notation $\mathcal{D}=(x_1,...,x_k,...,x_N)$ may be used directly without specifying the source and destination nodes, and $|\mathcal{D}|$ can be used in place of $N$.

For a given link $N_i\!\xrightarrow{l}\!N_j$, an ML model can be trained, given the usage statistics of the links. For this reason, the dataset $\mathcal{D}$ was split into two disjoint datasets $\mathcal{D}_{\mathrm{tr}}$ and $\mathcal{D}_{\mathrm{te}}$, with training and test data, respectively.

Let $x_k \in \mathcal{D}_{\mathrm{tr}}$ be the current sample, the less recent $n_{\mathrm{p}}$ samples (i.e., $X_k= (x_k, x_{k-1},...,x_{k-n_{\mathrm{p}}+1})$) are used to train a model $f_M(\theta^*, \cdot)$ and to estimate its hyperparameters $\theta^*$ for the prediction of the next target $x_{k+1}$ sample (i.e., if the slot associated to the link will be used for transmission or not). The trained ML model $f_M(\theta^*, X_k)$ is then used for the test dataset $\mathcal{D}_{\mathrm{te}}$ to predict $x_{k+1}$.
As usual, the goal for the training procedure is to minimize the mean squared error (MSE) between the prediction and the target $x_{k+1}$:

\begin{equation}
\theta^* = \arg \min_{\theta} \sum_{k=n_{\mathrm{p}},...,|\mathcal{D}_{\mathrm{tr}}|-1}\Big( x_{k+1} -f_M(\theta, X_k)\Big) ^2,
\end{equation}\smallskip

where $|\mathcal{D}_{\mathrm{tr}}|$ is the cardinality of the training dataset.

The main objective of this work is to study how the performance provided by ML models varies as the level changes in the case of periodic traffic patterns. For links related to leaf nodes, where $l=1$, the occupation of specific slots by the traffic is quite predictable, while at higher levels, as approaching the root node, flows belonging to the underlying nodes are increasingly mixed, and they become less and less predictable.

From an application perspective, a model $f^l_M(\theta^*, \cdot)$ could be trained for each link. In this article, we focus on how performance varies at different levels of the network topology. Since in the experimental campaign we selected only one link for each level, the information of the link on which the model is training is omitted in order to make the notation simpler.

As the objective of this paper was not to detect the best ML model but to analyze a generic model at different network levels, the very simple multilayer perceptron (MLP) model was selected. In addition, MLP can be implemented easily and efficiently, even in nodes with limited computational power. The MLP models exploited in this work consist of a two-layer neural network with $n_{\mathrm{p}}=890$ inputs, a hidden layer composed of 8 neurons with \textit{ReLu} activation functions, and one output layer with a \textit{sigmoid} activation function. The input vector $X_k$ has a size of  $n_{\mathrm{p}}=890$ corresponding to about \SI{30}{minutes}, considering that the schedule of the TSCH matrix repeats over time every $T_{\mathrm{matrix}}=\SI{2.02}{s}$. The model was trained in $20$ epochs using the \textit{adam} optimizer and the MSE as loss function. The \textit{batch size} was set equal to $32$, and the learning rate was initialized to $0.01$ and halved at each training epoch. The \textit{Keras} module included in \textit{TensorFlow} was used to manage both the training and the test phases of the model.

In addition, using the same test datasets $\mathcal{D}_{\mathrm{te}}$ used to assess the accuracy of the $f^l_M(\theta^*, \cdot)$ models, a statistical index that would give an indication of the difficulty in predicting the target at different levels has been computed.

In particular, to provide insight into the predictability of binary data, the normalized discrete autocorrelation was used:

\begin{eqnarray}
\rho_k & = & \frac{R_k}{R_0} , \ \ \ k \in  \left[-\left\lfloor\frac{|\mathcal{D}_{\mathrm{te}}|}{2}\right\rfloor,...,\left\lfloor\frac{|\mathcal{D}_{\mathrm{te}}|}{2}\right\rfloor \right],\\[\baselineskip]
R_k & = & \sum_{n=1}^{|\mathcal{D}_{\mathrm{te}}|} x_n \cdot x_{n+k},
\end{eqnarray}\smallskip

where $R_k$ is the non-normalized discrete autocorrelation.

High values of $\rho_k$ indicate that the vector corresponding to $\mathcal{D}_{\mathrm{te}}$ and the same vector translated by a quantity $k$ are strongly related, and consequently, their content is more predictable. On the contrary, low autocorrelation suggests a weak or no linear dependence with previous values, and consequently, data that is difficult to predict. By definition, $\rho_0=1$ and $\rho_k=\rho_{-k}$.

In particular, the evolution of the maximum value of the autocorrelation $\rho^l_{\mathrm{max}}$ of links at different network levels $l$ is analyzed in the next result section:

\begin{equation}
\rho^l_{\mathrm{max}} = \max_{k \ne 0} \rho_k.
\end{equation}\smallskip

\section{Results}
\label{sec:results}

The experimental campaign was carried out on the network reported in Fig.~\ref{fig:net}, which is a 4-layer network composed of $31$ nodes. Packet flows are generated by leaf nodes, numbered from $N_1$ to $N_{16}$, and the generation periods are reported in the rectangles below the nodes. The path for which results are reported and discussed is highlighted with a dashed red line. However, experimental results for the other paths are similar and are not reported here for space reasons.

\subsection{Dataset acquisition}

\begin{figure}[t]
\begin{center}
\includegraphics[width=1.0\linewidth]{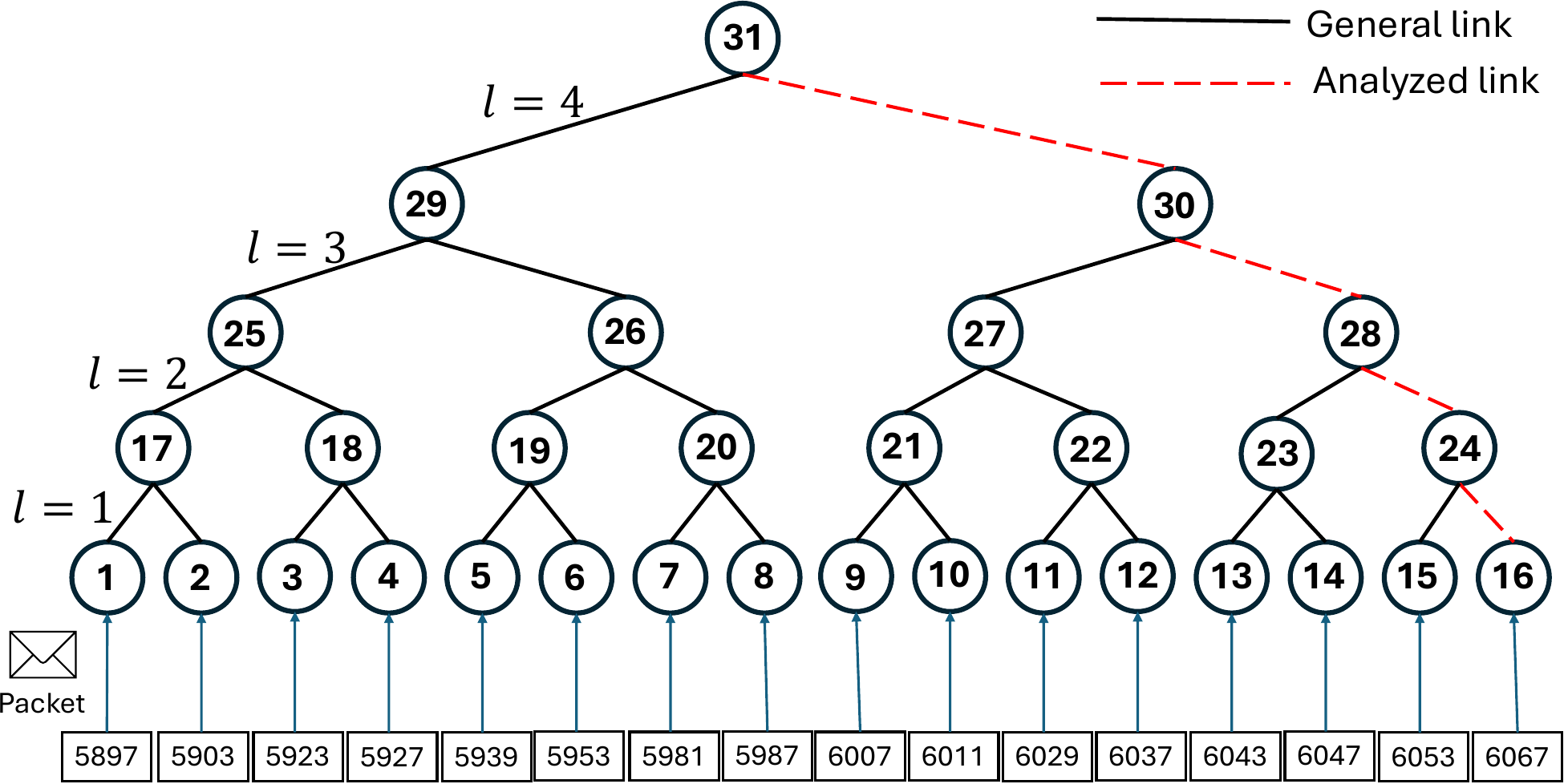}
\end{center}
\caption{Topology of the network considered in the experimental analysis.\label{fig:net}}
\end{figure}

The dataset was acquired by means of the TSCHmodeler discrete event simulator, which was already used in~\cite{10710912}.
An initial experimental campaign was conducted (details not included here for brevity) that demonstrates how the results from the simulator are very similar to those obtained from actual TSCH implementation based on OpenMote B devices running OpenWSN protocol stack (6TiSCH).

\begin{table}[t]
  \caption{\vspace{\mymargin}Main simulation parameters.}
  \label{tab:parameters}
  \small
   
  \begin{center}
    \tabcolsep=0.18cm
    \def\arraystretch{1.15}
    \begin{tabular}{l|l|c}
    \toprule
    Quantity & Description & Typical \\
     &  & value \\
    \midrule
    $N_\mathrm{slot}$  & Number of slots in a TSCH matrix & 101 \\
    $T_\mathrm{slot}$  & Duration of a slot & \SI{20}{ms} \\
    $T_\mathrm{matrix}$  & Duration of the TSCH matrix & \SI{2.02}{s} \\
    $N_\mathrm{try}$ & Maximum number of tries & 16 \\
    $\epsilon_\mathrm{frame}$  & Data frame success probability & 0.874 \\
    $\epsilon_\mathrm{ACK}$  & ACK frame success probability & 0.92 \\
    $T_\mathrm{sim}$ & Duration of the simulation & \SI{20}{years} \\
    \midrule
    $E_\mathrm{tx}$ & Energy to transmit data and receive ACK & $\SI{266}{\mu J}$ \\
    $E_\mathrm{rx}$ & Energy to receive data and transmit ACK  & $\SI{284}{\mu J}$ \\
    $E_\mathrm{listen}$ & Energy for idle listening (wasted) & $\SI{138}{\mu J}$ \\
    \bottomrule
    \end{tabular}
  \end{center}
\end{table}

The main simulation parameters are summarized in Table~\ref{tab:parameters}. Concerning the number of slots in the TSCH matrix, the duration of the single slot, the duration of the TSCH matrix, and the maximum number of allowed transmission attempts, we set the default values used by OpenWSN for OpenMode B nodes, i.e., $N_\mathrm{slot}=101$, $T_\mathrm{slot}=\SI{20}{ms}$, $T_\mathrm{matrix}=\SI{2.02}{s}$, $N_\mathrm{try}=16$.

The value of the data and ACK frames success probability (i.e., $\epsilon_\mathrm{frame}=0.874$ and $\epsilon_\mathrm{ack}=0.92$, respectively) were obtained from~\cite{9903301}, and they are based on data derived from a real testbed based on the OpenWSN implementation of the TSCH protocol.
The length of the simulation was set equal to \SI{20}{years}. The first \SI{12611881}{} samples were used for $\mathcal{D}_{\mathrm{tr}}$, while the latter \SI{3000000}{} samples were used for $\mathcal{D}_{\mathrm{te}}$ (i.e., $|\mathcal{D}_{\mathrm{tr}}|=\SI{12611881}{}$ and $|\mathcal{D}_{\mathrm{te}}|=\SI{3000000}{}$ samples).

Regarding power consumption, the measurements described in~\cite{9187609} for OpenMote B devices installed with OpenWSN were used as reference. It is interesting to notice that the energy to transmit a data frame of maximum dimension $E_{\mathrm{tx}}=\SI{266}{\mu J}$ (and receive the related ACK frame) or to receive a data frame of the same maximum dimension $E_{\mathrm{rx}}=\SI{284}{\mu J}$ (and to send the related ACK frame) are almost the same. Instead, the energy wasted due to idle listening $E_{\mathrm{listen}}=\SI{138}{\mu J}$ is almost half $E_{\mathrm{tx}}$ and $E_{\mathrm{rx}}$. This means that the waste of energy due to idle listening is quite high, considering the fact that for this type of network, most of the scheduled slots are typically unused. Other devices are characterized by different consumption. For instance, OpenMoteSTM devices have the following values in terms of power consumption, $E_{\mathrm{tx}}=\SI{485.7}{\mu J}$, $E_{\mathrm{rx}}=\SI{651.0}{\mu J}$, $E_{\mathrm{listen}}=\SI{303.3}{\mu J}$ \cite{2014-SJ-consumption}, but in any case, the waste of energy related to idle listening remains significant.

\begin{table}[t]
  \caption{\vspace{\mymargin}Configuration of the TSCH matrix used in the experimental campaign (in \textbf{bold}, analyzed links).\label{tab:matrix}}
  \small
   
  \begin{center}
    \tabcolsep=0.18cm
    \def\arraystretch{1.15}
    \begin{tabular}{|c|c|c|c|c|c|c|}
    \cmidrule{1-3}\cmidrule{5-7}
    Slot & Source & Dest. & & Slot & Source & Dest. \\
    Offset & Node & Node & \ \ \ \ & Offset & Node & Node \\
    \cmidrule{1-3}\cmidrule{5-7}
9 & 11 & 22 & & 45 & 22 & 27 \\
13 & 1 & 17 & & \textbf{46} & \textbf{24} & \textbf{28} \\
15 & 18 & 25 & & \textbf{50} & \textbf{16} & \textbf{24} \\
19 & 26 & 29 & & 57 & 13 & 23 \\
20 & 27 & 30 & & 59 & 15 & 24 \\
\textbf{21} & \textbf{30} & \textbf{31} & & 66 & 25 & 29 \\
28 & 20 & 26 & & 67 & 6 & 19 \\
31 & 7 & 20 & & \textbf{70} & \textbf{28} & \textbf{30} \\
32 & 9 & 21 & & 78 & 2 & 17 \\
33 & 19 & 26 & & 82 & 23 & 28 \\
34 & 8 & 20 & & 83 & 5 & 19 \\
35 & 29 & 31 & & 84 & 12 & 22 \\
37 & 10 & 21 & & 85 & 17 & 25 \\
38 & 14 & 23 & & 89 & 3 & 18 \\
40 & 21 & 27 & & 96 & 4 & 18 \\

    \cmidrule{1-3}\cmidrule{5-7}
    \end{tabular}
  \end{center}
\end{table}

The configuration of the TSCH matrix used in the experimental campaign was reported in Table~\ref{tab:matrix}, in which the slot offset used for every link of the network in Fig.~\ref{fig:net} was selected randomly.
The channel offset was not reported since we set $\epsilon_\mathrm{frame}$ and $\epsilon_\mathrm{ACK}$ equal for all channels.

In Table~\ref{tab:matrix} and in Fig.~\ref{fig:net}, the links $N_{16}\!\xrightarrow{l=1}\!N_{24}$, $N_{24}\!\xrightarrow{l=2}\!N_{28}$, $N_{28}\!\xrightarrow{l=3}\!N_{30}$, $N_{30}\!\xrightarrow{l=4}\!N_{31}$ are highlighted since they are those for which different performance metrics have been reported and analyzed.

In the experimental setup proposed in this work, traffic flows are generated by leaf nodes, one for each leaf.
Traffic is periodic, as is typical for this kind of network. The generation period for each flow was intentionally chosen to be different from the other flows. This decision was made to reflect the fact that in typical real networks, packet generation by applications is not synchronized with the network slots. In particular, the generation period measured in terms of time slot was reported in the lower part of Fig.~\ref{fig:net}. 

\subsection{Autocorrelation and Prediction Quality}

\begin{table}[b]
\centering
\tabcolsep=0.15cm
\def\arraystretch{1.15}
\small
\caption{\vspace{\mymargin}Confusion matrix and performance metrics at different levels\label{tab:results}}

\begin{tabular}{l | c c c c}
\toprule
Quantity & $l=1$ & $l=2$ & $l=3$ & $l=4$  \\
\midrule
True Positive (TP)  & 48444   & 92363   & 167481  & 312544 \\ 
False Negative (FN) & 13631   & 32017   & 81728   & 186919 \\
False Positive (FP) &     0   & 17034   & 52643   & 163289 \\
True Negative (TN)  & 2937035 & 2857696 & 2697258 & 2336358 \\
\bottomrule
\end{tabular}

\vspace{0.5cm}

\begin{tabular}{c | c | c c c c c}
\toprule
Link/Level & $\rho^l_{\mathrm{max}}$ & Acc. & Prec. & Rec. & F1 score & AUC \\
\midrule
$N_{16}\!\xrightarrow{l=1}\!N_{24}$ & 0.986 & 0.995 & 1.000 & 0.780 & 0.877 & 0.998 \\
$N_{24}\!\xrightarrow{l=2}\!N_{28}$ & 0.943 & 0.984 & 0.844 & 0.743 & 0.790 & 0.992 \\
$N_{28}\!\xrightarrow{l=3}\!N_{30}$ & 0.855 & 0.955 & 0.761 & 0.672 & 0.714 & 0.976 \\
$N_{30}\!\xrightarrow{l=4}\!N_{31}$ & 0.678 & 0.883 & 0.657 & 0.626 & 0.641 & 0.926 \\
\bottomrule
\end{tabular}
\end{table}

The datasets obtained for the four links at different levels were used to train the corresponding models $f^{l=1}_M(\cdot)$, $f^{l=2}_M(\cdot)$, $f^{l=3}_M(\cdot)$, and $f^{l=4}_M(\cdot)$.

The results evaluated on the test datasets  $\mathcal{D}_{\mathrm{te}}$ for the four links are reported in Tables~\ref{tab:results}.
The upper Table~\ref{tab:results} reports the values of the confusion matrices for different levels, namely: \textit{true positive} (TP, i.e., slots actually used that are predicted as used), \textit{false negative} (FN, i.e., slots actually used that are predicted as not used), \textit{false positive} (FP, i.e., slots actually not used that are predicted as used), and \textit{true negative} (TN, i.e., slots actually not used that are predicted as not used).

Regarding the lower Table~\ref{tab:results}, the first column represents the maximum normalized discrete autocorrelation $\rho^l_{\mathrm{max}}$. As expected, it is inversely proportional to the level $l$, and it reflects the growing complexity of the prediction task as levels increase. As previously mentioned, this occurs since traffic flows generated by nodes at the leaf level mix together in an unpredictable way.

The other five columns represent, in order, the \textit{accuracy} (i.e., how often the prediction of $f_M(\theta^*,\cdot)$ is correct), the \textit{precision} (i.e., the percentage of predicted used slots that are actually used), the \textit{recall} or \textit{sensitivity} (i.e., how many actual used slots are correctly predicted), the \textit{F1 scores} (i.e., a harmonic mean of precision and recall), and the \textit{AUC} that is an index ranging from $1.0$ (i.e., perfect prediction) to $0.5$ (i.e., random prediction). All these performance indicators can be computed directly from the confusion matrices reported in the upper table.

As highlighted by the table, all indices, in a different way, reflect a substantial decrease in performance as the level increases.

\begin{figure}[t]
\begin{center}
\includegraphics[width=1\linewidth]{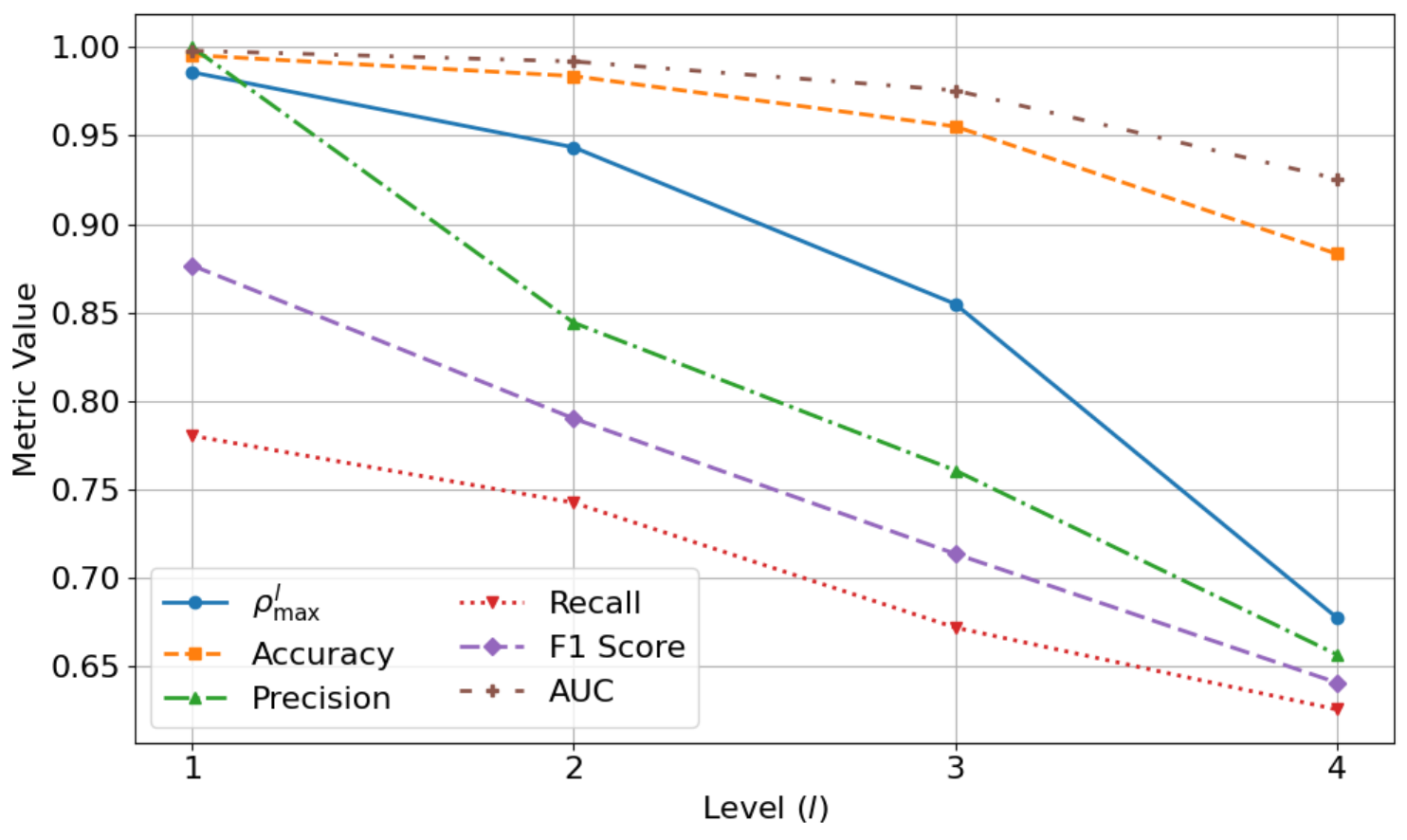}
\end{center}\
\caption{The evolution of key quality indicators as levels change.}
\label{fig:indices}
\end{figure}

The plot of Fig.~\ref{fig:indices} reports the evolution of the quality indices reported in Table~\ref{tab:results} as the level increases. As can be noticed, the most affected indices are precision and F1 score, since they have a sharper decreasing slope.

Regardless of the level, all the models exhibited a reasonably good prediction quality. In particular, the accuracy index shows that slots predicted as used have a probability of $88.3\%$ (in the most complex condition regarding $l=4$) of being effectively used by nodes. Unfortunately, as highlighted by the recall index, there are some slots that are predicted as not used for transmission, for which a transmission occurs. In this case, the transmission is delayed to the next available slot associated with the link, which is predicted as usable by the model. This increases the transmission latency, which may be acceptable for many application contexts of this type of network, where, in many cases, the main goal is to optimize power consumption.

This investigation is aimed at verifying the ability of ML models to predict slot usage at different levels of a TSCH/TDMA network. For the integration of the mechanism in a real TSCH network, some small modifications to the protocol are needed. In particular, since the ML model is intentionally not aware of the number of messages in the queue of the sender node, it might predict that a slot is not usable even if it has some frames ready for transmission in its transmission queue. 

To avoid this problem, if the transmission queue contains, in addition to the packet that is currently transmitted, another packet addressed to the same destination node (i.e., related to the same link), the transmitter has to add to the current transmitted packet an indication that the queue is not empty. In this way, even if the ML model predicts that the slot will not be used, the receiver will remain active in reception when the same link is scheduled for the next time. To this extent, the TSCH protocol provides specific \textit{information elements}, which are user-defined information that can be included in the MAC header. In the context of this work, the information element is added to notify the receiver if the queue is not empty.

In the case of loss of a data packet, it is retransmitted in the next slot, which is marked as usable by the ML algorithm. If the queue starts to grow, meaning that more frames have arrived in the meantime after the packet to be retransmitted, the system adapts to prevent the queue from increasing indefinitely. As described, this operation is performed by including an information element in the transmitted packet that disables the ML algorithm responsible for predicting slot usage.

\subsection{Energy consumption}

The main target of the proposed analysis is the use of ML algorithms for traffic prediction aimed at reducing power consumption.

Even if the exact computation of the energy saved can be performed only by means of a complete implementation of the proposed algorithm, including the optimizations based on information elements described in the previous subsection, some considerations can be derived directly from the results obtained in this context.

\begin{table}[t]
\centering
\tabcolsep=0.18cm
\def\arraystretch{1.15}
\small
\caption{\vspace{\mymargin}Energy consumption of a specific link vs. level}
\label{tab:energy_res}
\begin{tabular}{l | c c c c}
\toprule
Quantity & $l=1$ & $l=2$ & $l=3$ & $l=4$  \\
\midrule
$P_{\mathrm{tx}}$ ($\SI{}{\mu W}$)    &  2.73 & 5.46 & 10.94 & 21.92 \\
$P_{\mathrm{rx}}$ ($\SI{}{\mu W}$)    &  2.91 & 5.83 & 11.68 & 23.41  \\
$P_{\mathrm{listen}}$ ($\SI{}{\mu W}$)                   & 0.00 & 0.39 & 1.20 & 3.72   \\
$P^{\neg \mathrm{ML}}_{\mathrm{listen}} ($\SI{}{\mu W}$)$ & 66.88 & 65.46 & 62.62 & 56.92 \\
\bottomrule
\end{tabular}
\end{table}

Specifically, the value reported in Table~\ref{tab:energy_res}, are optained from the confusion matrices reported on Table~\ref{tab:results}.
The number of transmissions and related receptions can be approximated as $\mathrm{TP}+\mathrm{FN}$, where FN refers to the slots that should be exploited for transmission but the ML algorithm has chosen not to utilize. It is not an error to count them in the computation of energy consumption, since these transmissions are performed in any case, but using different slots. For the exchange of messages, referred to a specific link, the power consumed by the node is:

\begin{eqnarray}
P_{tx} & = & \frac{(\mathrm{TP}+\mathrm{FN})\cdot E_{\mathrm{tx}}}{|\mathcal{D}_{\mathrm{te}}|\cdot T_{\mathrm{matrix}}}, \\[\baselineskip]
P_{rx} & = & \frac{(\mathrm{TP}+\mathrm{FN})\cdot E_{\mathrm{rx}}}{|\mathcal{D}_{\mathrm{te}}|\cdot T_{\mathrm{matrix}}},
\end{eqnarray}\smallskip

where the denominator $|\mathcal{D}_{\mathrm{te}}|\cdot T_{\mathrm{matrix}}$ represents the duration in seconds of the test dataset $\mathcal{D}_{\mathrm{te}}$.

The mechanism proposed in this work acts by preventing idle listening by not turning on the node in reception if the ML algorithm does not predict the usage of the slot.
Consequently, TN represents those slots for which the node is set in a sleep state since they are not used for transmission, while FP represents those slots that suffer from idle listening because their usage is not properly predicted by the ML algorithm.
The additional power wasted for idle listening is:

\begin{equation}
P_{\mathrm{listen}} = \frac{n_{\mathrm{listen}}\cdot E_{\mathrm{listen}}}{|\mathcal{D}_{\mathrm{te}}|\cdot T_{\mathrm{matrix}}},
\label{eq:idle}
\end{equation}\smallskip

where $n_{\mathrm{listen}}=\mathrm{FP}$. If in (\ref{eq:idle}) we substitute $n_{\mathrm{listen}}=\mathrm{FP}+\mathrm{TN}$, we obtain $P^{\neg \mathrm{ML}}_{\mathrm{listen}}$, which represents the energy wasted for idle listening for a typical TSCH implementation without the use of ML to predict slots utilization.

As highlighted by the results reported in Table~\ref{tab:energy_res}, the component of idle listening, which is predominant in the communication, is nevertheless completely eliminated in practice. In the worst operating condition, corresponding to $l=4$, $P_{\mathrm{listen}}$ is reduced to $\SI{3.72}{\mu W}$, which has to be compared with the unmodified TSCH implementation where $P^{\neg \mathrm{ML}}_{\mathrm{listen}}=\SI{56.92}{\mu W}$. As expected, for links closer to the root node, the capability of the ML model to predict slot usage decreases, and the energy waste slightly increases.

\section{Conclusions}
\label{sec:conclusions}

This paper investigated the performance of machine learning algorithms to capture the traffic patterns generated by the TSCH communication protocol for WSNs, at different distances from the source nodes in multi-hop networks.
Its final goal is the use of the inferred information to reduce the energy usage of the sensor nodes by turning the nodes in deep sleep mode when they are not actively involved in the communication.

Experimental results show the effectiveness of the proposed approach while highlighting the negative impact of the network topology in terms of hops on the performance of the algorithm.
These results open several interesting future perspectives, including the definition of all the details to integrate the mechanism in the TSCH protocol, and the experimentation of different network topologies, machine learning algorithms, and traffic patterns. 

The main limitation, and at the same time source of inspiration for future work, is that this paper does not account for the extra energy required for machine learning algorithms to run on the node.
Additionally, only cyclic generation patterns were analyzed, which are nevertheless the most common in the application context of this kind of WSNs.

\bibliographystyle{IEEEtran}
\bibliography{bibliography}

\end{document}